\documentclass[mathleft
]{an}
\usepackage{graphicx}
\usepackage{times}
\usepackage{psfig}
\usepackage{float}
\usepackage{psfrag}
\usepackage{amsmath}
\usepackage{amssymb}
\usepackage{ulem}
\usepackage{ctable}
\usepackage{txfonts}
\usepackage{enumerate}

\newcommand{\ysz}{$Y_{\mathrm{SZ}} ~$}
\newcommand{\plk}{{\it Planck~}}

\begin{document}

% The following seven commands are intended for editorial usage and should be ignored by
% the author(s).
\Pagespan{789}{}% Document's page range. 
% If second parameter is left empty, the last page is computed automatically.
\Yearpublication{2012}%
\Yearsubmission{2012}%
\Month{}%   
\Volume{333}%  
\Issue{}% 
% \DOI{This.is/not.aDOI}% 

%\title{Radio-SZ correlation for known radio halo clusters in the {\it Planck} \\ Early Sunyaev-Zel'dovich catalog}
%\title{Some results on the correlation between cluster radio halo power \\ and the Sunyaev-Zel'dovich effect signal}
\title{Some results on the radio-SZ correlation for galaxy cluster radio halos} 

\author{Kaustuv Basu\thanks{\email{kbasu@astro.uni-bonn.de}\newline}
%Example 
%for footnote, note the usage of the \texttt{fnmsep}
%command as separator between institute number and footnote mark} 
}
\titlerunning{An SZ take on radio halos}
\authorrunning{K. Basu}
\institute{
Argelander Institut f\"ur Astronomie, Universit\"at Bonn, 
D-53121 Bonn, Germany
}

\received{3 Sep 2012}
\accepted{17 Sep 2012}
\publonline{later}


\abstract{%
We present correlation results for the radio halo power in galaxy clusters with the integrated thermal Sunyaev-Zel'dovich (SZ) effect signal, including new results obtained at sub-GHz frequencies. The radio data is compiled from several published works, and the SZ measurements are taken from the \plk ESZ cluster catalog. The tight correlation between the radio halo power and the SZ effect demonstrates a clear correspondence between the thermal and non-thermal electron populations in the intra-cluster medium, as already have been shown in X-ray based studies. The radio power varies roughly as the square of the global SZ signal, but when the SZ signal is scaled  to within the radio halo radius the correlation becomes approximately linear, with reduced intrinsic scatter. We do not find any strong indication of a bi-modal division in the radio halo cluster population, as has been reported in the literature, which suggests that such duality could be an artifact of X-ray selection. We compare the \ysz dependence of radio halos with simplified predictions from theoretical models, and discuss some implications and shortcomings of the present work.
}
\maketitle

\section{Introduction}
Precision use of  galaxy clusters for measuring the cosmological parameters largely depends on our ability to accurately describe the properties of the intra-cluster medium (ICM). The ICM contains about 90\% of the baryonic mass in massive clusters, and reaches very high temperatures ($\sim$2-15 keV) after attaining thermal equilibrium in the deep cluster potential well. This thermal plasma emits in the X-rays through the bremsstrahlung emission, and is also visible in the radio/sub-millimeter wavebands through the thermal Sunyaev-Zel'dovich (SZ) effect: the two methods that have been highly successful in finding galaxy clusters from large-area surveys and studying their detailed properties.  

In addition to this dominant thermal component the ICM also hosts a rich variety of non-thermal energy sources and magnetic fields. The most spectacular evidence for this cluster-wide non-thermal phenomena comes from the observation of giant radio halos, which are diffuse synchrotron emission  extending over $\sim 1$ Mpc scales. They are presumed rare, typically found in dynamically active systems (mergers), and have a smooth distribution similar to the ICM that indicates a connection between their powering mechanism and the cluster mass (e.g. Liang et al. 2000, Ferrari et al. 2008). What remains unknown, however, is the detailed mechanism by which these GeV electrons get distributed in the cluster volume, and whether their presence or absence mark special sub-populations of clusters not suitable for precision cosmology. We also do not know if the radio halo power can be used as an accurate and unbiased mass indicator in merging systems, where the standard mass estimation method based on hydrostatic equilibrium is obviously not adequate.

With the goal of establishing a cleaner mass selection in the radio halo studies, we obtained the first radio-SZ correlation in galaxy clusters (Basu 2012). The radio data were collected from the literature, and for the SZ measurements we used the \plk Early SZ cluster catalog (Planck collaboration 2011). While we cannot answer questions regarding the rate of occurrence of radio halos in a given mass bin since we lack  radio observation of a mass-limited sample, our work aims to provide answers to the mass scaling of radio halos and the scatter in their radio power. In particular, we have attempted to find out if the observed {\it bi-modal distribution} between radio halo and radio ``quiet" clusters, as seen from  X-ray selected samples (e.g. Brunetti et al. 2007), is an inherent property of radio halos, or merely a selection artifact. This bi-modality in the radio halo cluster population has been used as an argument in support of radio halo origin models based on turbulent re-acceleration of electrons following cluster mergers (Brunetti et al. 2007 and references therein).

\begin{figure*}
\centering
\includegraphics[width=8.4cm]{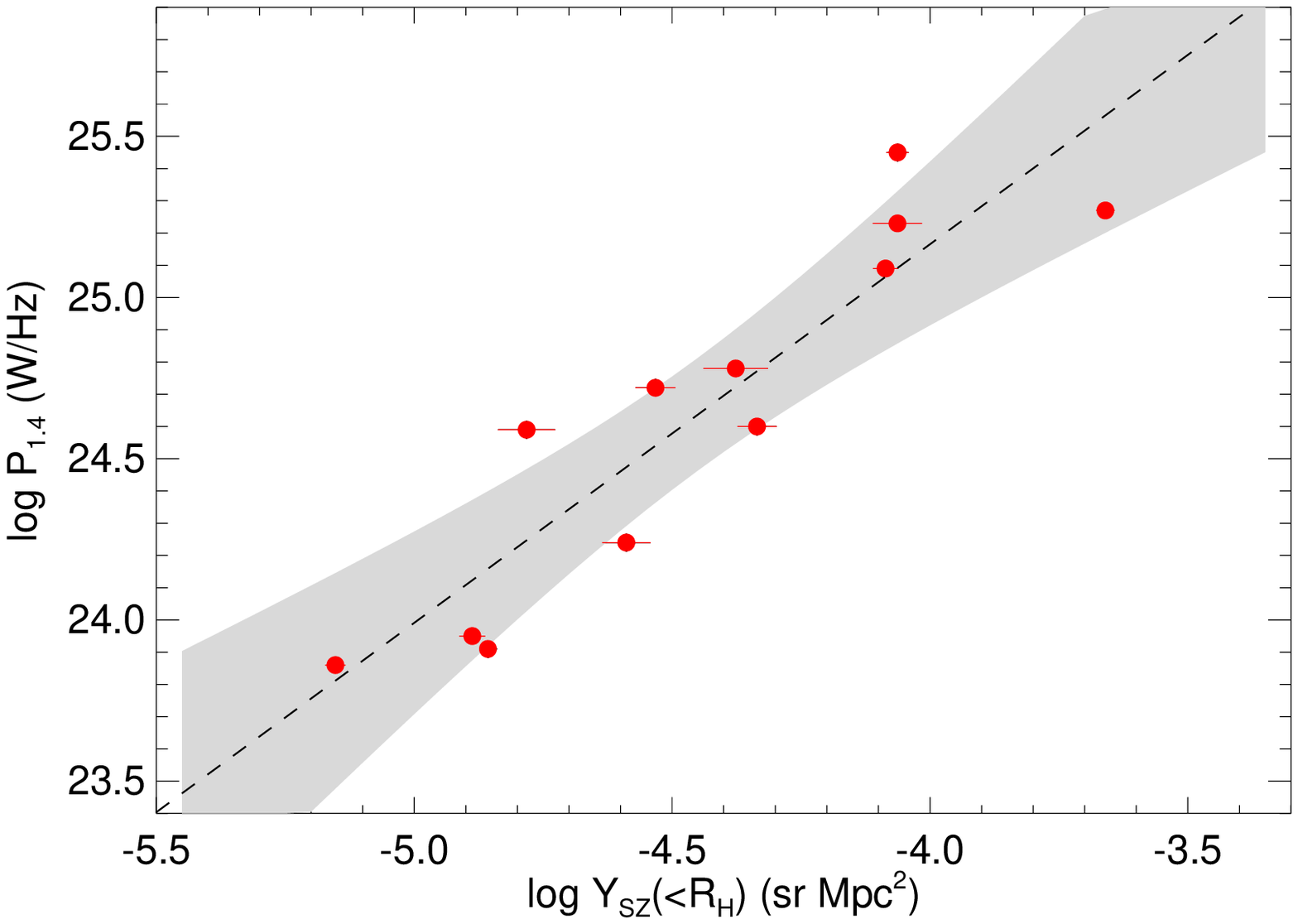}%
\includegraphics[width=8.4cm]{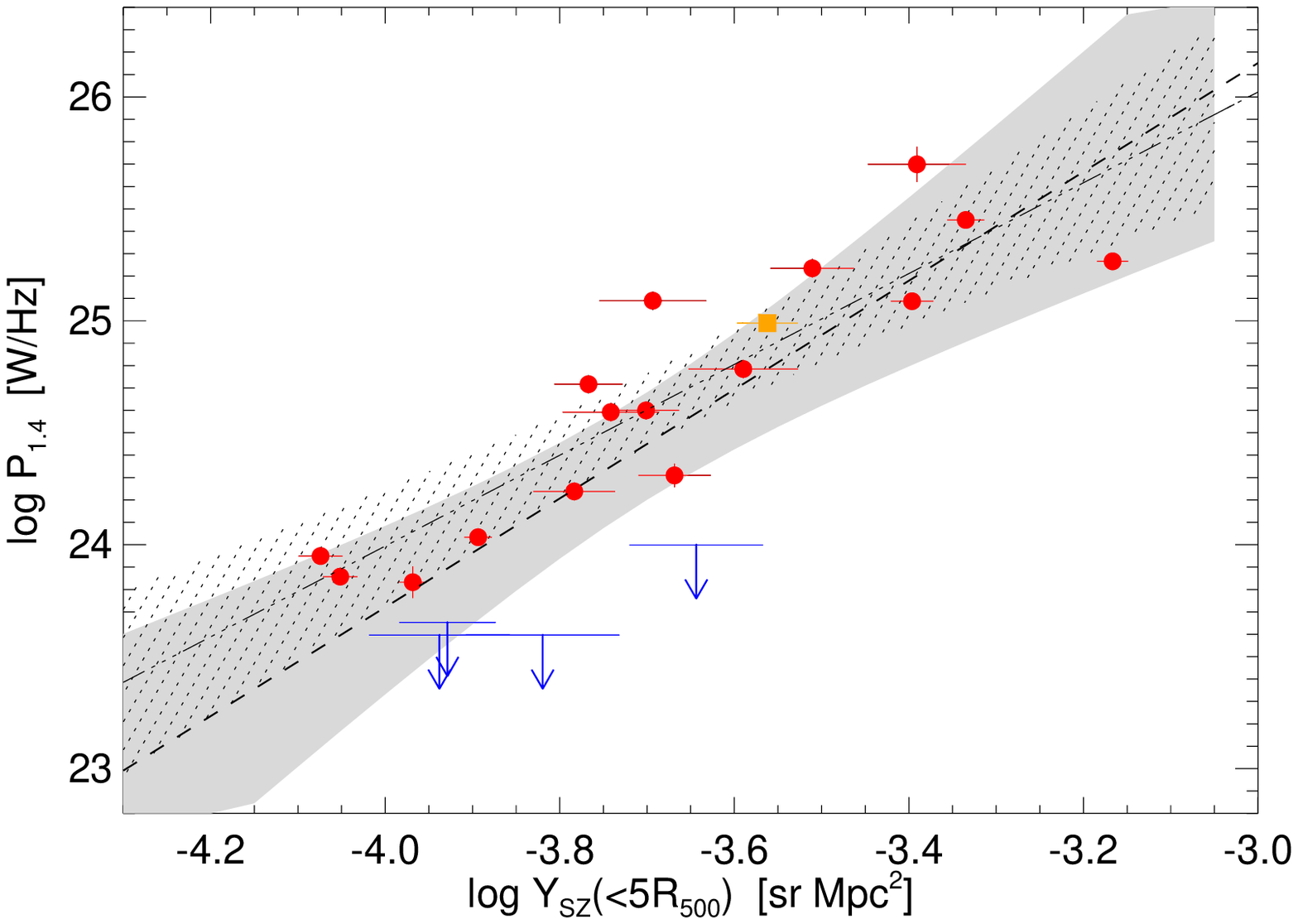}
\caption{{\it Left:} $P_{1.4}-Y_{\mathrm{SZ}}$ correlation for a sample of clusters with radio halos as compiled by Cassano et al. (2007). The integrated SZ signal from the \plk catalog have been scaled to inside the halo radius assuming the ``universal" GNFW pressure profile. The slope of the scaling is slightly super-linear, with intrinsic radio scatter roughly 0.28 dex. 
{\it Right:} $P_{1.4}-Y_{\mathrm{SZ}}$ correlation and testing of a bi-modal division in the radio halo population using the Brunetti et al. (2009) sample. Here \ysz represents the total SZ signal inside radius $5R_{500}$ (as given in the \plk catalog), and the slope is steeper compared to the scaling inside radio halos. The red circles denote the radio halo detections, blue arrows the non-detection upper limits (see  Brunetti et al. 2007) and the orange squares is a mini halo (A2390). The short dashed line represents the fit for halos only, and the long dashed line when upper limits are included. The shaded regions mark the 95\% confidence intervals for the corresponding fits.
}
\label{correl}
\end{figure*}

%A robust correlation between the radio halo power and the SZ signal can also be expected based on the timescale argument: the rapid boost in the X-ray luminosity during mergers happens in a relatively short timescale, compared to a more gradual increase in the integrated SZ signal with moderate fluctuations (Poole et al. 2007, Wik et al. 2008). This delayed increment of the SZ signal should correspond better with radio halo timescales ($\sim 1$ Gyr), derived from the spatial extent of the radio halos. The SZ signal is also a more robust indicator of cluster mass than the X-ray luminosity, irrespective of cluster dynamical state or other details of cluster physics (e.g. Motl et al. 2005, Reid \& Spergel 2006). Thus the SZ-selection might be able to show the presence of radio haloes in late mergers and other massive clusters which are potentially left out in X-ray luminosity limited samples.

Compared to X-ray luminosities, the integrated SZ signal is a more robust indicator of cluster mass, having little dependence on the cluster dynamical state or other details of cluster physics (e.g. Motl et al. 2005, Reid \& Spergel 2006). During cluster mergers both $L_{\mathrm X}$ and \ysz go through transient variabilities, but the changes in \ysz are less extreme than in $L_{\mathrm X}$ (Poole et al. 2007, Wik et al. 2008), thus less affecting the mass selection. The gradual late-time increase in the SZ signal should also correspond better with the radio halo timescales ($\sim 1$ Gyr, derived from the spatial extent of the radio halos). As a result, we can expect the SZ selection to reveal the presence of radio haloes in late mergers and other massive clusters which are potentially left out in X-ray luminosity limited samples.

\section{Results for the radio-SZ correlation}

In the following we present some highlights of our work using the \plk ESZ clusters that have known radio halo data in the literature. All results are derived using the $\Lambda$CDM concordance cosmological parameters. The quantity \ysz denotes the intrinsic Compton $Y$-parameter, $Y d_A^2 E(z)^{-2/3}$, where $d_A$ is the angular distance and $E(z)$ is the ratio of the Hubble parameter at redshift $z$ to its present value. The \plk catalog provides the integrated $Y$ parameter within the radius $5R_{500}$, obtained from a multi-frequency matched filtering (see Planck collaboration 2011).  This is much larger than the typical radio halo extension, so we also obtained a {\it scaled} \ysz value inside individual radio halo radius using the universal pressure profile of Arnaud et al. (2010). Both these total and scaled integrated SZ signals are then correlated with the radio halo power.

\subsection{\ysz and mass scaling of radio halos}

For radio catalogs we used the compilations from Giovannini et al. (2009) and Cassano et al. (2007), which give radio halo power measured at 1.4 GHz. The latter sample is used due to its better definition of radio halo radius. The correlation of the radio power with the total SZ signal ($Y_{\mathrm{SZ}}(<5R_{500})$, as measured by \plk) follows a roughly {\it quadratic} power-law: $P_{\nu} \propto Y_{\mathrm{SZ}}^{1.8 \pm 0.4}$ for the Giovannini et al. (2009) sample, and $P_{\nu} \propto Y_{\mathrm{SZ}}^{1.9 \pm 0.2}$ for the Cassano et al. (2007) sample. Scaling the SZ signal inside halo radius makes the correlation approximately linear: e.g. for the Cassano et al. (2007) sample we obtain $P_{\nu} \propto Y_{\mathrm{SZ}}(<R_H)^{1.2 \pm 0.2}$, with intrinsic scatter 0.24 dex (Fig.\ref{correl} {\it left}). We use this slightly super-linear relation for obtaining the mass scaling for radio halos, although we emphasize from the current data a linear relation between radio and SZ power inside halos is a fully valid result. This linear relation is  expected to become sub-linear at lower frequencies for curved spectrum of radio halos.

The relation between the integrated SZ signal and cluster mass is $Y_{\mathrm{SZ}} \propto f_{\mathrm{gas}} M_{\mathrm{tot}}^{5/3} E(z)^{2/3}$, which has been found to be observationally robust (e.g. Andersson et al. 2011, Planck collaboration 2012). $f_{\mathrm{gas}}$ is the gas mass fraction which varies weakly with the cluster mass in the high-mass regime: $f_{\mathrm{gas}} \propto M_{500}^{0.14}$ (e.g. Bonamente et al. 2008). Using the same gas fraction at all radii we therefore obtain $P_{\nu} \propto M_{\mathrm{tot}}^{3.4 \pm 0.4}$ for the total virial mass, or $P_{\nu} \propto M_H^{2.1 \pm 0.3}$ for the mass enclosed within a radio halo. 

The mass scaling of radio halos within the halo radius is consistent with previous estimates using X-ray hydrostatic masses (e.g. Cassano et al. 2007). The global scaling with $M_{500}$ or $M_{\mathrm{vir}}$ is steeper than that determined from X-ray luminosities (e.g. Cassano, Brunetti \& Setti 2006), and can be considered more robust due to \ysz being a better mass proxy. Directly applying the $L_X-M_{500}$ scaling to our samples, we obtain $P_{\nu} \propto M_{500}^{2.4}$, which is consistent with the mass scaling inside radio halos, but much flatter than the scaling with total virial mass. This is likely due to the very peaked nature of the X-ray luminosity profile, causing most of the X-ray emission originating from within $\sim r_{2500}$ (which is similar to the radio halo radius), and therefore correlating more directly with the physical state of the ICM inside radio halo radius. 

\begin{figure*}
\centering
\includegraphics[width=8.4cm]{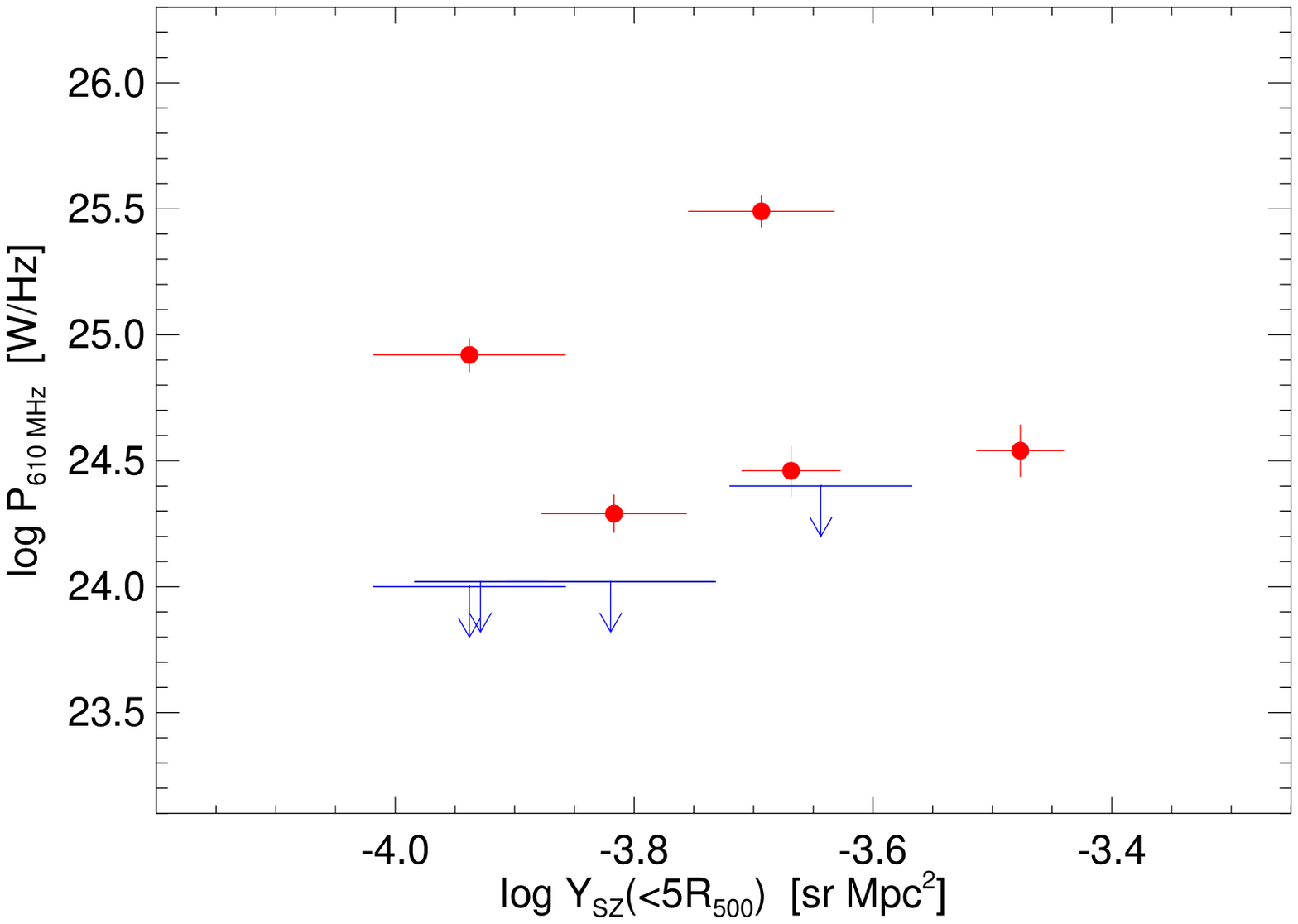}
\includegraphics[width=8.4cm]{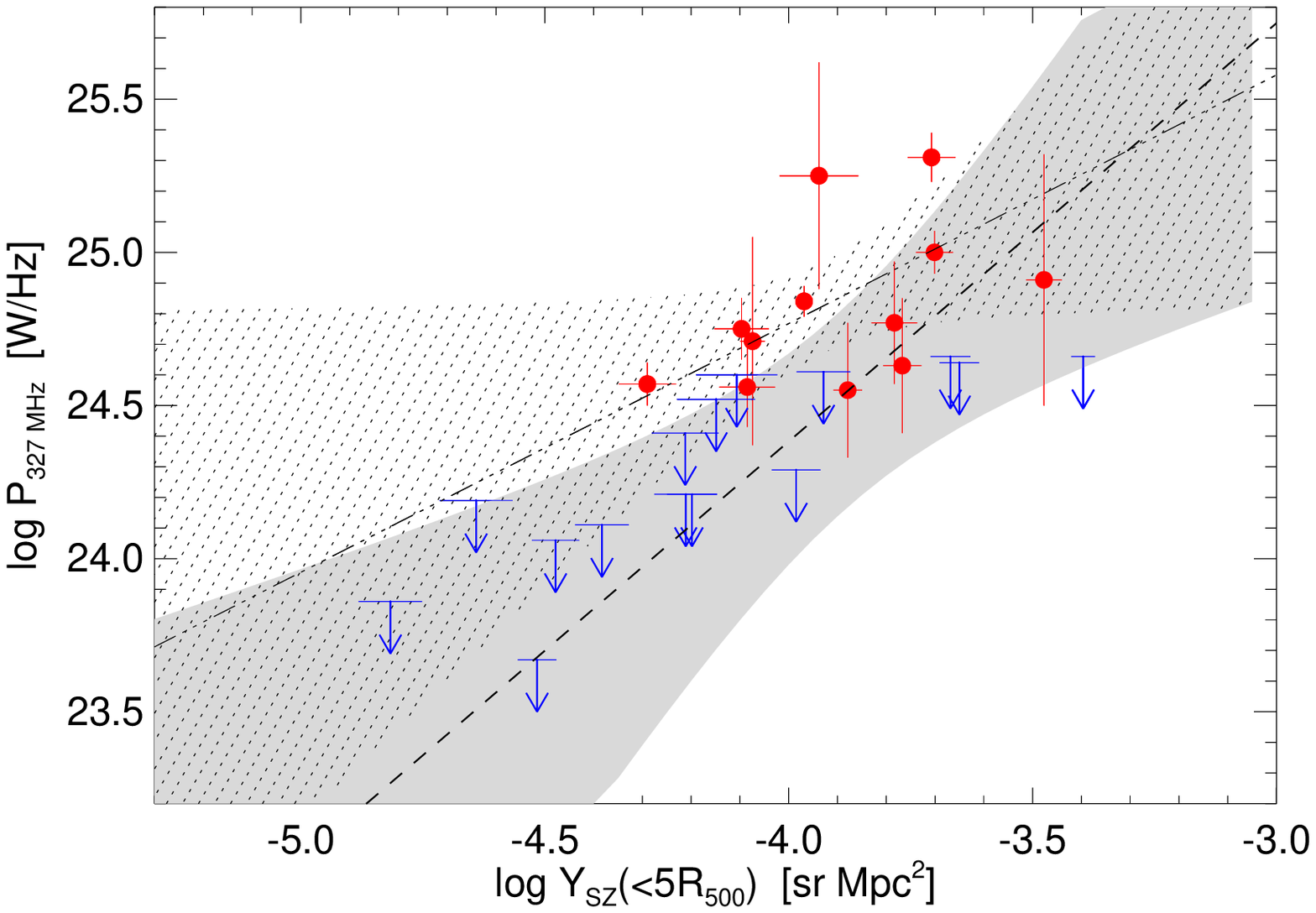}
\caption{{\it Left:} Radio-SZ correlation result at 610 MHz from the cluster sample observed by Venturi et al. (2008). This study is based on 26 clusters selected above $L_X > 5\times 10^{44}$ erg/s, but the corresponding number of counterparts in the \plk catalog is too few to obtain any robust SZ-radio correlation. This plot highlights the intrinsic scatter in the radio halo power, although flux under-estimation in the GMRT data can be an issue. 
{\it Right:} Radio-SZ correlation at 327 MHz from the sample of Rudnick \& Lemmerman (2009), based on WENSS radio data. The survey data is shallow, and correspondingly the non-detection upper limits are not sufficiently below the detection line to test a bi-modality. 
%The data nevertheless indicates a flattening of the radio-SZ correlation slope at this low frequency. 
All data points and lines have the same meaning as in Fig.\ref{correl}. 
}
\label{lowfreq}
\end{figure*}

\subsection{The issue of bi-modality}

For testing whether there are two distinct populations of clusters, namely those hosting powerful radio halos and those without, we use radio halo catalogs that include non-detections. A  systematic study of radio halo non-detections in an X-ray selected sample is done by Venturi et al. (2008), using GMRT observation at 610 GHz. However, this sample contains too few clusters which have \plk SZ measurements to obtain any robust fit (Fig.\ref{lowfreq} {\it left}). We therefore use the compilation by Brunetti et al. (2009), which lists the same GMRT results scaled to 1.4 GHz with a list of other halo detections. In their work, non-detection upper limits were obtained by simulating fake radio halos in the GMRT data and scaling to 1.4 GHz by using $\alpha=1.3$ for the spectral index. The result of the correlation is shown in Fig.\ref{correl} {\it right}. There appears to be a hint of bi-modal division, but the number of radio halo non-detection clusters in the \plk catalog is too low to confirm this with certainty. The point to note here is, we do not see high-\ysz objects with radio halo non-detections, as in the case of highly X-ray luminous radio ``quiet'' cool core clusters. A similar result is obtained when the radio halo power is correlated against the total gas mass inside $R_{500}$, by dividing the SZ signal with the mean temperature: $M_{\mathrm{gas}} = Y_{\mathrm{SZ}}(R_{500})/T_X ~~ \mu_e m_p / (\sigma_T/m_e c^2)$. We therefore conclude that the strong bi-modal division observed in the X-ray selected cluster samples is likely due to the inclusion of X-ray bright cool core clusters with lower masses, which do not show extended radio halos.

An interesting comparison between the X-ray and SZ selections can be obtained if we compare the relative rate of occurrence of radio halo clusters in the original X-ray selected samples with that in the \plk catalog (Table 1). Both the samples of Venturi et al. (2008) and Rudnick \& Lemmerman (2009), which are complete above certain X-ray luminosity thresholds, have radio halo fraction of roughly 20\%. This fraction increases to roughly 50\% when we look for the counterparts in the \plk catalog for these samples. We emphasize that this is {\it not} an objective test of radio halo statistics from SZ selection, since none of these samples are selected {\it a priori} based on their SZ signal, hence they are not complete above a mass threshold. Nevertheless, the point to highlight here is that the fraction of radio halo clusters is clearly increasing in an SZ-based sample, as opposed an X-ray luminosity limited samples. The most likely explanation is that \ysz is a better mass proxy and the \plk ESZ catalog is specifically showing the most massive clusters in the universe which are likely to harbor a significant non-thermal electron population.

%\vspace*{-0.3cm}
\begin{table}[width=\columnwidth]
\begin{center}
\caption{Relative frequency of finding radio halos in different X-ray selected cluster catalogs, and the corresponding rate of occurrence in the \plk ESZ catalog.}
 \vspace{2mm}

\begin{tabular}{|l||c|c||c|c|}
\hline
Sample & no. of & of which & \plk  & of which \\
  &  clusters &  radio halos  &  detects  &  radio halos  \\
  \hline\hline
  
 V08  &  26  &  6   &   9   &  5   \\ 
 B09$^{\star}$  &  44  &  21 &  20  &  16 \\
 R09  &  72  &  14 &  27  &  12 \\
 \hline
\end{tabular}
 \begin{minipage}[b]{0.96\columnwidth}
 \centering
 \vspace{1mm}
 Samples: V08=Venturi et al. 2008; B09=Brunetti et al. 2009; R09=Rudnick \& Lemmerman 2009. 
 $^{\star}$ Not X-ray complete
 \end{minipage}
\end{center}
\end{table}%

\vspace*{-0.3cm}
\subsection{Correlation at low frequencies}

The question of radio halo statistics and their mass scaling at sub-GHz frequencies is highly relevant for the ongoing and future radio surveys like LOFAR and SKA. We tried to obtain similar radio-SZ correlations for low-frequency radio halo observations of X-ray complete samples, but the current data sets do not allow robust constraints. Venturi et al. (2008) sample is based on the GMRT observation at 610 MHz of a X-ray selected sample. Out of 26 clusters only 9 are present in the \plk catalog, which does not allow a robust regression analysis, but the result highlights the scatter in the radio halo power and inter-mixing of halos and non-detections (Fig.\ref{lowfreq} {\it left}). We remark, however, that inadequate sampling of the short baselines in the GMRT 610 MHz measurements can result in serious under-estimation of radio halo fluxes (e.g. Dallacasa et al. 2009).

Rudnick \& Lemmerman (2009) re-analyzed the WENSS survey data at 327 MHz (Rengelink et al. 1997) for an X-ray selected sample. The shallowness of the WENSS data, which is roughly an order of magnitude less sensitive than typical VLA radio halo observations scaled to this frequency, does not allow a comprehensive test for the bi-modality. However, the method used by these authors to extract radio halo fluxes, based on simulating fake halos in empty fields, safeguards against the flux loss issue. 
%Their simulations show lack of any flux under-estimation bias at the $\geq 3\sigma$ detection significance used for their catalog.  
Correlation of the 327 MHz radio halo power with the global \ysz signal (Fig.\ref{lowfreq} {\it right}) yields a slope much shallower than those observed in the 1.4 GHz samples when only $\geq 3\sigma$ halo detection are considered: $P_{0.33} \propto  Y_{\mathrm{SZ}}^{0.81\pm 0.35}$.  
The large statistical errors resulting from the shallow WENSS data hides the intrinsic scatter in the radio power, which dominates the 1.4 GHz radio uncertainties. 
When non-detections are included with the measured values, the slope becomes compatible with the 1.4 GHz result at roughly 1.2$\sigma$, but the scatter also increases to almost twice as much. %($P_{0.33} \propto  Y_{\mathrm{SZ}}^{1.38\pm 0.43}$). 
%The large statistical errors resulting from the shallow WENSS data hides the intrinsic scatter in the radio power, which dominates the 1.4 GHz radio uncertainties.

The turbulent re-acceleration model of radio halo origin predicts a synchrotron power-law spectrum that gradually steepens at higher frequencies (e.g. Brunetti \& Lazarian 2011). Based on such steepening one can naively expect the radio-SZ correlation to become flatter at lower frequencies, as the radio under-luminous objects at 1.4 GHz will have a larger relative gain in their flux; although Cassano (2010) predicts a steepening of the $P_{\nu}-L_X$ correlation at low frequencies based on a model population of ultra-steep spectrum radio halos in low-mass clusters. The correlation result from Rudnick \& Lemmerman (2009) shows no clear evidence of steepening, and is at best consistent with our 1.4 GHz results if we consider the halos and no-detections as one population. The gradually steepening spectrum of radio halos should also lead to a reduced intrinsic scatter in the radio power at low frequencies, which unfortunately can not be tested from these data sets. 

\vspace*{-0.2cm}
\section{Discussion}

The observed linear correlation between the radio and SZ power inside radio halos can be provisionally explained using simplified hadronic models, which predicts $P_{\nu}  / Y_{\mathrm{SZ}} \propto X_{\mathrm{CR}} ~n ~\sigma_{\mathrm{pp}} ~f_B$ (e.g. Kushnir et al. 2009). Here $X_{\mathrm{CR}}$ is the ratio between cosmic ray pressure and thermal pressure, $n$ is the gas density, $\sigma_{\mathrm{pp}}$ is the $p$-$p$ collision cross-section, and $f_B$ is the volume filling factor for magnetic fields. This simple formulation requires constant cosmic ray proton density and large magnetic field strength over the entire radio halo volume, which may be unrealistic. The turbulent re-acceleration of particles for creating radio halos requires a complex mechanism, and again we use a simplified model based on the works by Cassano \& Brunetti (2005) and Cassano et al. (2007). these authors obtain a scaling $P_{\nu}  \propto Y_H T_e^{1/2}$, which is in good agreement with the observed mild super-linear correlation. However, the analytic relationship between the radio halo power and the cluster mass and temperature in this model is based on several simplifying assumptions, and the parameters require fine tuning to match observations. The true explanation of the observed quasi-linear correlation between the radio and SZ signals can therefore be beyond these simplified models.

The small number of non-detections in the \plk ESZ catalog does not allow us to conclude whether the bi-modality seen in the X-ray selected samples is absent -- or merely weaker -- when measured against the cluster SZ signal. Likewise, we have not been able to address the true fraction of radio halos in clusters in a given mass range, since the current work is based on prior X-ray selected clusters that are present in the \plk catalog. Some of these questions will be addressed in a future publication utilizing public radio survey data and complete cluster samples in the SZ and X-ray (M. Sommer \& K. Basu, in preparation).
%This issue will also be addressed in the future works with radio survey data as well as new observations.

%\section{Summary}

%In this article we have summarized our recent findings on the radio-SZ correlation of galaxy clusters harboring radio halos, with additional details on the scaling results at low frequencies.
\medskip
\acknowledgements
The author thanks the organizers of the XMM-Newton 2012 science workshop in Madrid for providing the opportunity to present this work.

%%%%%%%%%%%%%%%%%%%%%%%%%%%%%%%%%%%%%%%%%%%%%%%%%%%%%%

\end{document}